\documentclass{sig-alternate}
\usepackage{graphicx}
\usepackage{tabularx}
\usepackage{multirow}
\usepackage{balance}
\usepackage{times}
\usepackage{array}
\usepackage{amsmath}
\usepackage{enumitem}
\newcolumntype{L}[1]{>{\raggedright\let\newline\\\arraybackslash\hspace{0pt}}m{#1}}
\newcolumntype{C}[1]{>{\centering\let\newline\\\arraybackslash\hspace{0pt}}m{#1}}
\newcolumntype{R}[1]{>{\raggedleft\let\newline\\\arraybackslash\hspace{0pt}}m{#1}}

\usepackage{graphicx}
\usepackage{caption}
\usepackage{subcaption}

\conferenceinfo{HT}{2015 Cyprus, Turkey}

\def\sharedaffiliation{%
\end{tabular}
\begin{tabular}{c}}
\begin{document}
\title{Language, Twitter and Academic Conferences}

\numberofauthors{5} 
\author{
\alignauthor Ruth Garc\'ia$^1$
\alignauthor Diego G\'omez$^2$
\alignauthor Denis Parra$^2$
\and
\alignauthor Christoph Trattner$^3$
\alignauthor Andreas  Kaltenbrunner$^1$
\alignauthor Eduardo Graells-Garrido$^4$
\sharedaffiliation
\affaddr{$^1$Barcelona Media. Barcelona, Spain}\\
\affaddr{$^2$Pontificia Universidad Cat\'olica de Chile. Santiago, Chile}\\
\affaddr{$^3$NTNU. Trondheim, Norway}\\
\affaddr{$^4$Telef\'onica I+D. Santiago, Chile}\\
}

\maketitle

\begin{abstract}

Using Twitter during academic conferences is a way of engaging and connecting an audience inherently multicultural by the nature of scientific collaboration. English is expected to be the \emph{lingua franca} bridging the communication and integration between native speakers of different mother tongues. However, little research has been done to support this assumption. In this paper we analyzed how integrated language communities are by analyzing the scholars' tweets used in 26 Computer Science conferences over a time span of five years. We found that although English is the most popular language used to tweet during conferences, a significant proportion of people also tweet in other languages. In addition, people who tweet solely in English interact mostly within the same group (English monolinguals), while people who speak other languages tend to show a more diverse interaction with other \emph{lingua groups}. Finally, we also found that the people who interact with other Twitter users show a more diverse language distribution, while people who do not interact mostly post tweets in a single language. These results suggest a relation between the number of languages a user speaks, which can affect the interaction dynamics of online communities.





\end{abstract}



\section{Introduction}

In the past few years, Twitter has been used as a conference backchannel platform in academic events targeting the expansion of the community communication and participation 
\cite{ebner2009introducing,ross2011enabled}. Attendees using Twitter are generally involved in note taking, sharing resources and reporting individual real-time reactions to events, covering both conference presentations and conference social activities. 
This supports scholars' activities such as disseminating their work and engaging general public and newcomer scientists into the research communities \cite{mitchell2007places}. It is a common practice in research conferences to use hashtags in the tweets to identify that particular event (e.g. \#hypertext2015).  
International academic conferences have a diverse community, with different cultural backgrounds and languages. Thus, it is interesting to analyze how language affects the generation of content and interaction among attendees. Such study would allow to observe how integrated a research community is, as well as to identify its blind spots in communication.
This can be of special interest to conference organizers not only to evaluate communication but also to have an overview of their audiences. Despite the research done in the past \cite{dyl43, Letierce2010,WenCSCW14,WenHT14} on academic conferences, little has been done on language communities and the communication established among them. 
To bridge this gap, we explore the language of 7M tweets posted by 18K users during 26 Computer Science conferences over five years (one week before and after for each conference). 
We group users by the language(s) they use to tweet in order to explore how different language communities interact.
Although English is expected to be the lingua franca of many international events, we wonder to what extent people use other languages on Twitter during academic conferences. 

\textbf{Research Questions.} Overall, our study was driven by the following research questions:

\begin{itemize}[itemsep=0pt,parsep=0pt,topsep=0pt, partopsep=0pt]

\item \textbf{RQ1. Conference attendees' languages}: To what extent do people tweet in other languages beyond English in conferences? 


\item \textbf{RQ2. Interactions between lingua groups}: How do lingua groups interact with each other? 


\item \textbf{RQ3. Effect of language}: Is there an effect of language or lingua group over online user interaction?






\end{itemize}


\textbf{Main results.} We find that most people tweet only in English (61\%)  in conferences but most of the tweets are posted by multilingual users and their participation varies significantly across conferences. 

Additionally, we observe that \emph{English monolinguals} receive most of the attention and interact more within their group while the opposite is observed with most of the members from other language communities.
Finally, we show that people who do not interact other attendees are mostly monolinguals, while people who interact with others present more language diversity, by a balanced distribution of monolinguals and multilinguals.  



 
\vspace{-3mm}
\section{Dataset} \label{dataCollections}

We selected a representative set of conferences in Computer and Information Science from the CORE Computer Science Conference Ranking list\footnote{http://www.core.edu.au/index.php/conference-rankings}; 26 conferences active in Twitter every year between 2009 and 2013. Furthermore, we manually checked that the selected conferences did not overlap with other events. To retrieve the tweets from these events in previous years, we used the Topsy API and crawled tweets containing the corresponding official hashtag (e.g., \#chi12, \#www2009) within a two-week time window around the dates each conference took place (from seven days before and until seven days after the conference ended). We found that these tweets were posted by 22,021 participants in total. We acknowledge that these participants also interact with others without the conference hashtag and because of this we also crawled their timeline tweets during the same period. In total, we obtained \emph{6,993,693} tweets.

\textbf{Language Identification.} To identify the language of the tweets, we removed all URLs, mentions and hashtags. Then we set a minimum threshold of \emph{4} remaining words in the tweets to identify their language. The language detection task was performed with a professional language tool provided by Yahoo Labs Barcelona that is able to identify over 40+ languages as in \cite{Poblete2011}. Following this process we were left with 6,184,775 tweets (88\% from initial sample) with an identified language. Finally, we proceeded to model each user by the three most frequent languages they used to tweet (setting a minimum threshold of \emph{5} tweets per language). Consequently, we found \emph{266} lingua groups with \emph{18,347} users using at least three  different languages in their tweets. 



\section{Results}\label{RQs}

\begin{table}[t!]
\centering
\scalebox{0.8}{ 
\begin{tabular}{|l|r|r|r||r|r|r|r|}
\cline{5-8}
\multicolumn{4}{c||}{}
&\multicolumn{4}{c|}{\textbf{Diversity percentage}}
 \\ \hline
 
&\multicolumn{3}{c||}{Lingua groups} 
&\multicolumn{2}{c|}{General} 
&\multicolumn{2}{c|}{Reciprocated} \\ \hline 
Conference & \multicolumn{1}{l|}{1-ling} & \multicolumn{1}{l|}{2-ling} & \multicolumn{1}{l||}{$\geq$ 3-ling} 
& \multicolumn{1}{l|}{MT}
& \multicolumn{1}{l|}{RT}
& \multicolumn{1}{l|}{MT}
& \multicolumn{1}{l|}{RT} \\ \hline
AAAI    &     81\%    &     8\%    &     11\%    &    34\%   &    29\%   &    16\%   &    20\%\\ \hline
ACMMM    &     52\%    &     38\%    &     11\%    &    53\%   &    53\%   &    48\%   &    41\%\\ \hline
CHI    &     76\%    &     17\%    &     7\%    &    49\%   &    48\%   &    40\%   &    30\%\\ \hline
CIKM    &     66\%    &     24\%    &     10\%    &    54\%   &    54\%   &    44\%   &    40\%\\ \hline
ECIR    &     58\%    &     27\%    &     15\%    &    55\%   &    57\%   &    43\%   &    31\%\\ \hline
ECIS    &     57\%    &     31\%    &     12\%    &    46\%   &    44\%   &    24\%   &    0\%\\ \hline
HT    &     64\%    &     26\%    &     10\%    &    52\%   &    53\%   &    37\%   &    29\%\\ \hline
ICIS    &     67\%    &     26\%    &     8\%    &    44\%   &    41\%   &    19\%   &    16\%\\ \hline
ICML    &     75\%    &     17\%    &     8\%    &    52\%   &    55\%   &    20\%   &    21\%\\ \hline
ICMT    &     51\%    &     30\%    &     20\%    &    70\%   &    62\%   &    31\%   &    20\%\\ \hline
ICSE    &     58\%    &     32\%    &     10\%    &    47\%   &    46\%   &    40\%   &    47\%\\ \hline
ISMAR    &     64\%    &     28\%    &     8\%    &    39\%   &    37\%   &    19\%   &    21\%\\ \hline
IUI    &     62\%    &     21\%    &     17\%    &    59\%   &    58\%   &    45\%   &    44\%\\ \hline
KDD    &     73\%    &     18\%    &     10\%    &    53\%   &    50\%   &    38\%   &    37\%\\ \hline
MobileHCI    &     66\%    &     23\%    &     11\%    &    50\%   &    47\%   &    48\%   &    39\%\\ \hline
NIPS    &     74\%    &     19\%    &     8\%    &    46\%   &    48\%   &    25\%   &    20\%\\ \hline
SIGGRAPH    &     77\%    &     16\%    &     7\%    &    38\%   &    32\%   &    24\%   &    19\%\\ \hline
SIGIR    &     68\%    &     21\%    &     12\%    &    56\%   &    58\%   &    36\%   &    39\%\\ \hline
SIGMOD    &     72\%    &     23\%    &     6\%    &    58\%   &    53\%   &    19\%   &    12\%\\ \hline
SLE    &     59\%    &     32\%    &     9\%    &    58\%   &    58\%   &    40\%   &    40\%\\ \hline
UBICOMP    &     71\%    &     21\%    &     9\%    &    59\%   &    57\%   &    55\%   &    44\%\\ \hline
UIST    &     71\%    &     24\%    &     5\%    &    60\%   &    58\%   &    35\%   &    32\%\\ \hline
VLDB    &     67\%    &     26\%    &     7\%    &    56\%   &    53\%   &    29\%   &    21\%\\ \hline
WSDM    &     65\%    &     22\%    &     13\%    &    61\%   &    60\%   &    48\%   &    39\%\\ \hline
WWW    &     52\%    &     32\%    &     15\%    &    52\%   &    51\%   &    43\%   &    40\%\\ \hline
XP    &     58\%    &     35\%    &     7\%    &    53\%   &    52\%   &    51\%   &    54\%\\ \hline
\end{tabular}
}
  \vspace{-1mm}  
\caption{Percentage of monolinguals, bilinguals and multilinguals tweeting in each conference between 2009-2013 (col 2-4). Diversity percentage for different type of interactions (col 5-8) .}
\label{tab:tbl_languages_number_total}
\vspace{-3mm}
\end{table}

\setlength\tabcolsep{1mm}
\begin{table}[t!]
\centering
\scalebox{0.8}{ 
 \begin{tabular}{|l|r|r|r|}
 \hline
 Lingua & Users & Tweets & (tweets/user) \\ \hline
en & 61.31\% & 29.14\% & 179.50\\ \hline
en-fr & 6.46\% & 3.57\% & 208.79\\ \hline
en-es & 3.79\% & 2.39\% & 238.14\\ \hline
de-en & 2.18\% & 1.63\% & 281.89\\ \hline
en-nl & 2.15\% & 1.50\% & 263.54\\ \hline
fr & 2.00\% & 0.26\% & 49.05\\ \hline
en-ja & 1.92\% & 3.55\% & 696.92\\ \hline
en-es-pt & 1.62\% & 4.06\% & 944.93\\ \hline
en-pt & 1.44\% & 0.35\% & 92.65\\ \hline
en-it & 1.36\% & 1.56\% & 434.83\\ \hline
nl & 1.36\% & 0.16\% & 43.33\\ \hline
ja & 1.09\% & 1.09\% & 377.89\\ \hline
en-es-fr & 0.93\% & 8.89\% & 3609.91\\ \hline
ca-en-es & 0.79\% & 2.14\% & 1016.69\\ \hline
en-ko & 0.57\% & 0.51\% & 340.24\\ \hline
es & 0.52\% & 0.06\% & 42.92\\ \hline
Others&	10.52\% &39.14\% &  - \\\hline
\end{tabular} 
}
  \vspace{-1mm}  
\caption{Statistics of top lingua groups (more than 90 users). We show the percentage of users belonging to each \emph{lingua} (Users), the percentage of tweets (Tweets) and the engagement (tweets/user).}\label{tbl_stats_linguas}
\vspace{-1mm}
\end{table}




\textbf{RQ1. To what extent do people tweet in other languages beyond English  across conferences?} \label{RQ1}

As expected, we found that the majority of tweets are written in English (76\%). 
Nevertheless, due to the multicultural nature of conferences, there is a non-negligible 24\% of tweets in languages different than English (en), such as French (fr), Spanish (es), German (de) and Japanese (jp). Furthermore, we found in our dataset that many people post tweets in more than a single language. 

We quantify this observation in Table \ref{tab:tbl_languages_number_total} that shows the percentage of users who tweet in a single language (1-lingua), in two languages (2-lingua) or three or more ($\geq$ 3-lingua) in each conference. We observe that the percentage of people who tweet in two or more languages goes from close to  20\% (AAAI, SIGGRAPH) up to around 50\% (ACMM, ICMT, WWW) showing important differences among conferences in the distribution of users who tweet in one or more languages. 
Based on these results, rather than analyzing languages as isolated groups, we studied the lingua groups as communities of people who speak either one or more languages. Table \ref{tbl_stats_linguas} describes the top language communities by number of users. The table shows that the majority of users are classified as \emph{English monolinguals} (61\%) but interestingly only produce (29\%) of all tweets with a moderate engagement (only 179.5 tweets per user). 
In contrast, we see that users of multilingual groups are the most engaged (3609.9 tweets/user for en-es-fr, 1016.7 for ca-en-es, and 944.93 for en-es-pt).

These results lead us to further analyze specific lingua groups to unveil the interaction between language communities and their online behavior.

\setlength\tabcolsep{1mm}
\begin{table}[t]
\centering
\scalebox{0.8}{ 
 \begin{tabular}{ |l|r|r|l|r|r|}  \hline

\multicolumn{6}{|c|}{\textbf{General}} \\ \hline 
\multicolumn{3}{|C{2cm}|}{\textbf{Mentions (148,184)}} & \multicolumn{3}{C{2cm}|}{\textbf{Retweets (91,523)}} \\
\hline

Ling.&Att.& out-links& Ling.&Att.& out-links \\ \hline
en & 67\% &  37\% & en & 66\% & 37\% \\ \hline
en-fr & 7\% & 56\% & en-fr & 7\% & 54\%\\ \hline
de-en & 3\% & 74\% & de-en & 3\% & 78\% \\ \hline
en-es & 3\% & 79\% & en-es & 3\% & 80\% \\ \hline
en-ja & 2\% & 35\% & en-ja & 2\% & 42\% \\ \hline \hline
\multicolumn{6}{|c|}{\textbf{Reciprocated}} \\ \hline
\multicolumn{3}{|C{2cm}|}{\textbf{Mentions (25,956)}} & \multicolumn{3}{C{2cm}|}{\textbf{Retweets (6,496)}}  \\ \hline
Ling.&Att.& out-links& Ling.&Att.& out-links \\ \hline
en & 57\% & 48\%   & en & 51\% &  52\%\\ \hline
en-fr & 8\% & 52\% & en-fr & 8\% &  44\%\\ \hline
de-en & 4\% & 72\% & en-es & 5\% & 61\%\\ \hline
en-es & 4\% & 71\% & de-en & 4\% & 74\%\\ \hline
en-nl & 3\% & 71\% & en-nl & 3\% & 70\%\\ \hline
\end{tabular} }
  \vspace{-3mm}  
\caption{\emph{Most popular linguas}: lingua groups ordered by the attention they receive across all conferences. The \emph{out-link} column represents the percentage of interactions going to other lingua groups.}\label{tbl_mostpop_lang}
\vspace{-5mm}
\end{table}

\textbf{RQ2. How do lingua groups interact with each other?} \label{RQ2}



\begin{figure}[t!]
        \centering
\begin{subfigure}[b]{0.4\textwidth}
                    \centering
\scalebox{0.8}{
	 \includegraphics[width=\linewidth]{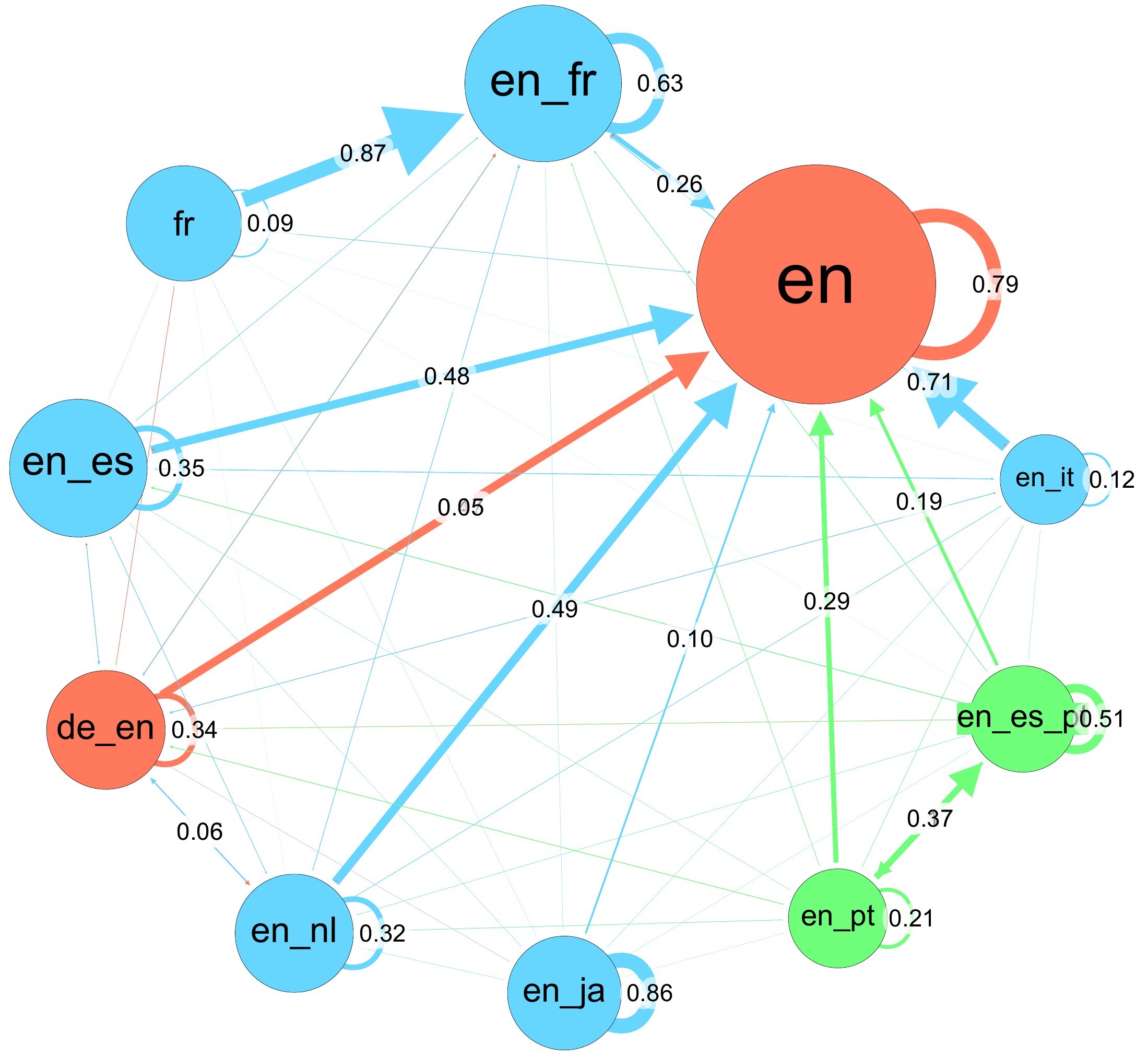} 	  	
	 }
\caption{Mentions between lingua groups. An edge from lingua group $x$ pointing to lingua group $y$ shows proportions of mentions that people in lingua group $x$ directed to people in lingua group $y$. For readability, we only show probabilities $\geq 0.05$.} 

\label{fig-top10}
\end{subfigure}\quad

\begin{subfigure}[b]{0.4\textwidth}
                    \centering
         \includegraphics[width=\linewidth]{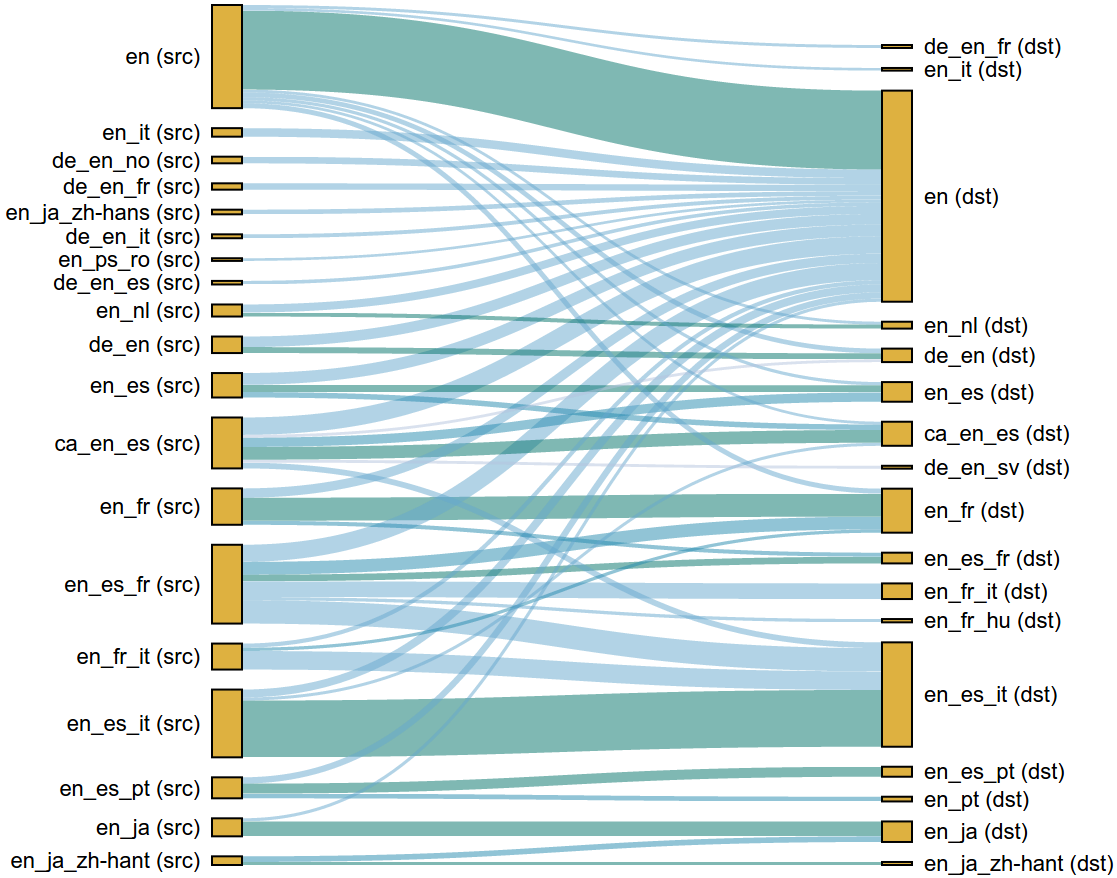} 	   		
\caption{Retweet interactions between top 50 most active lingua groups.}
\vspace{-3mm}
  		\label{fig-alllanguages}
\end{subfigure}  \quad     
  
\caption{\emph{(a)} Nodes representing the top 10 lingua groups based on mentions. \emph{(b)} Interactions between lingua groups based on source language (src) retweeting posts in a target language (dst).} \label{figlinguas}
  \vspace{-5mm}
\end{figure}


To answer this question, we first define two types of interactions: (1) general interactions and (2) reciprocated interactions. We refer to \emph{general interactions} to all retweets and also to tweets containing mentions, while \emph{reciprocated interactions} correspond to reciprocated retweets and tweets with mentions.    

Secondly, we measure diversity using the Gini-Simpson index, as in \cite{Chimera04,KimHT14} who called it \emph{diversity index}. This diversity index ranges from 0 to 1 and it measures the probability that two lingua groups taken at random from a set of interactions belong to different lingua groups. Participants of a conference with diversity index close to 0 will have the tendency to interact with people of the same lingua group. Conversely, conferences with values close to 1 show a uniform distribution of interactions with other lingua groups. We define diversity $D$ of a lingua group as: 
\vspace{-2mm}
 \begin{equation} 
D(c,i)= 1-\sum_{j\in S}\left({\frac{I_{i,j}^{c}}{N_{i}}}\right)^{2} 
\vspace{-3mm}
\end{equation}\label{eq:diversity}
with $N_{i}=\sum_{k\in S} {I_{i,k}^c}$  and where $I_{i,j}^{c}$ is the total number of interactions between people of lingua $i$ and $j$. $N_{i}$ is the total number of interactions of people of lingua $i$ in conference $c$. In order to know the diversity of a conference, we average $D(c,i)$ over all the linguas in conference $c$.  

We see in Table \ref{tab:tbl_languages_number_total} the diversity for each conference (we represented it as a percentage). We find some interesting patterns showing that a lower percentage of monolinguals is linked to higher diversity. For example, ICMT is  the most diverse conference for the general type of interactions and the percentage of monolinguals is the lowest of all (51\%). Conversely, AAAI shows high percentage of monolinguals (82\%) and the lowest diversity for the general interactions. On the other hand. reciprocal interactions do not show to be related to the percentage of monolinguals. For example, UBICOMP presents a high percentage of monolinguals and the highest diversity for the reciprocal interactions.

Furthermore, we look at the attention \emph{received} by members of each lingua by calculating  the number of mentions and retweets received from different users. Table \ref{tbl_mostpop_lang} shows the top 5 most popular lingua groups. Without doubt, English monolinguals are the most mentioned and retweeted in the general and reciprocated interactions. Albeit the fact that English monolinguals do not produce most of the tweets, they still receive most of the attention. This is mostly explained by the column \emph{out-links}, which shows the percentage of mentions and retweets about \emph{different} lingua group. For example, we see that only 37\% of the mentions and retweets generated by English monolinguals refer to other groups. Interestingly, Japanese bilinguals also prefer to interact mostly within their group. Conversely, groups like \emph{en-fr, de-en, en-es} refer more users of \emph{different} lingua groups in their interactions.

 
More evidence of the unequal activity between lingua groups is seen in Figure \ref{figlinguas}, which considers only the top 10 lingua groups and shows (a) the mentions network (general type) and (b) the retweet network (general type) across lingua groups. Figure \ref{fig-top10} shows that 79\% of all mentions from the \emph{en} group also belong to the same group. Moreover, 35\% of mentions from the \emph{en-es} lingua group refer to users from the same group, and 48\% to the \emph{en} group. 

In Figure \ref{fig-alllanguages}, the Sankey plot represents the network of retweets. Again, here we see that for most of the cases the English group retweets members from the same group. At the same time, the English group receives most of the attention from other language communities. Interestingly, in similar proportion, lingua groups en-es-it, en-fr, en-es-pt and en-ja show a similar pattern, preferably retweeting users on their same lingua groups.


\textbf{RQ3. Is there any effect of language or lingua group over online user interaction?} \label{RQ3}

\begin{figure}[t!]
\vspace{-4mm}
        \centering
\scalebox{0.8}{
	 \includegraphics[width=\linewidth]{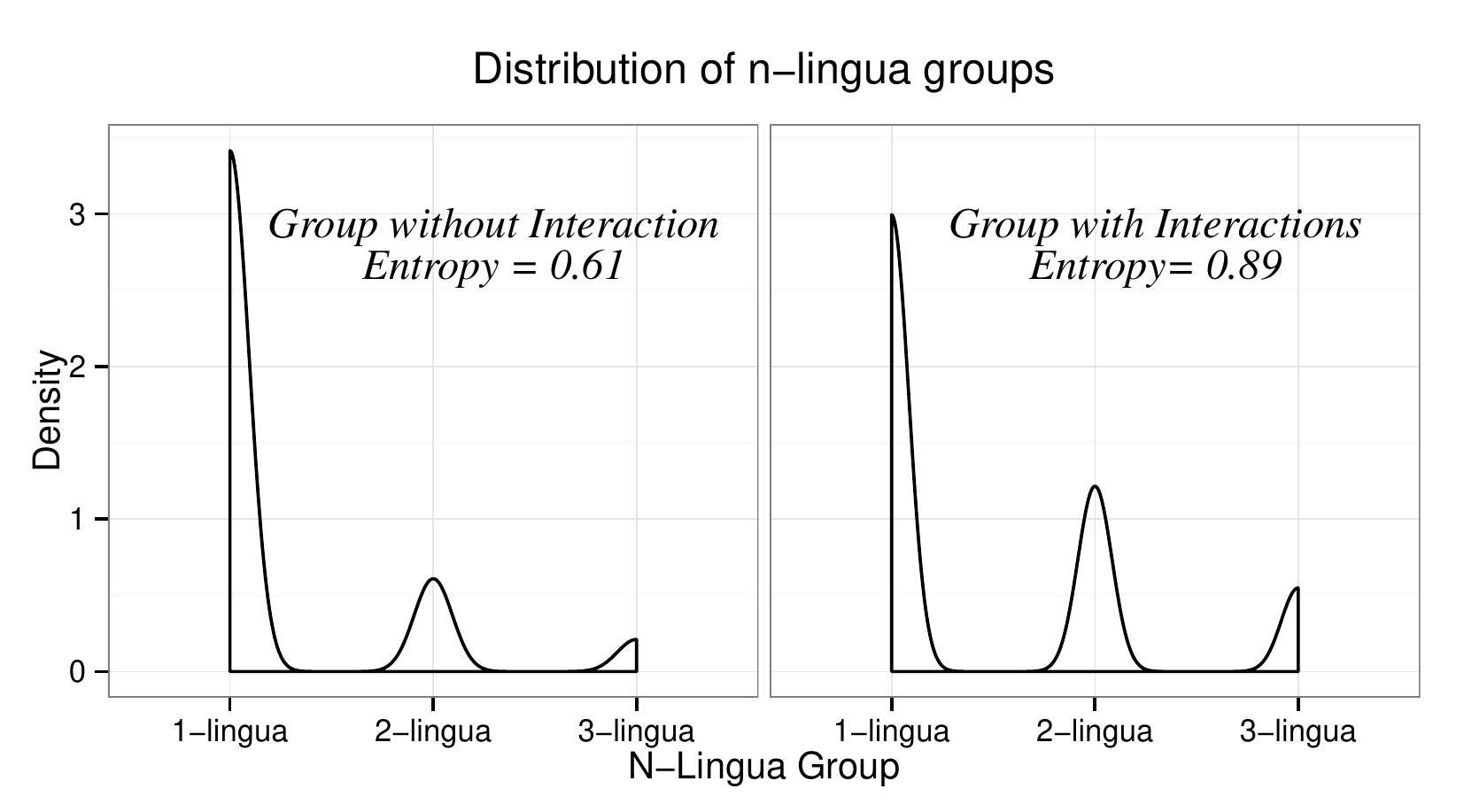}
}	 
	 \vspace{-4mm}
\caption{Distribution of n-lingua groups considering users without (left graph) and with reply/mention interactions (right graph).} 
\vspace{-4mm}
\label{fig-lingua-group}
\end{figure}

We addressed this question by studying how the number of languages a Twitter user speaks affects her online behavior. As already explained, if a user has posted tweets in only one language we consider her in the 1-lingua group (monolingual), while another user tweeting in two languages will be in the 2-lingua group, and so on. We found two results that show at general and at individual level the effect of the amount of languages on user interaction. At the general level, we found that among the users who posted tweets but who had not interacted with other people (by mentioning them), the percentage on monolinguals is considerably larger (80.6\%) than multilinguals. A different picture is seen among users who interacted at least once during the conference (by mentioning someone in a tweet), since only 62.9\% of those users are monolinguals and the rest are multilinguals. We conducted a chi-square test of proportions comparing the distribution of monolinguals, bilinguals and trilinguals between people who interacted and people who did not. We found a statistically significant difference with $\chi^2$ = 416.6, $df = 2$, $p\-value < .001$. This relation can be better observed in Figure \ref{fig-lingua-group}, where the group who interacted (right-side plot) had a more balanced distribution and hence a higher entropy (a measure of diversity \cite{shannon2001mathematical}) of $H(s) = 0.89$ compared to a smaller diversity on lingua groups among people who did not interact with an entropy $H(s) = 0.61$. 
Moreover, at the individual level we found that the more the languages a user speaks, the larger the likelihood to interact with others. Table \ref{tab-logit-results} shows the results of a logistic regression where the dependent variable measures whether the user \textit{interacted} with other people or not. The factors in the regression are the  \textit{year} of the conference and the number of languages the user has used to tweet (\textit{n\_languages}). We observe that the number of languages has a significant $\beta$ coefficient of 0.666 ($p<.001$), which can be interpreted by saying that, keeping all the other factors fixed, for each additional language the user speaks the odds ratio of interacting in the network increases by $95\%$ (since $e^{0.666} = 1.95$).

\begin{table}[!htbp] 
\small
  \vspace{-3mm}
\centering 
\begin{tabular}{@{\extracolsep{5pt}}l{c}{c}{c} } 
\\[-1.8ex]\hline 
\\[-1.8ex]  Variable  & $\beta$ coeff. & S.E. \\ 
\hline \\[-1.8ex] 
 year(=2009) & $2.049^{***}$ & $(0.390)$ \\ 
 year(=2010) & $2.458^{***}$ & $  (0.385)$\\ 
 year(=2011) & $2.453^{***}$ & $ (0.385)$ \\ 
 year(=2012) & $2.294^{***}$ & $ (0.383)$ \\ 
 year(=2013) & $2.423^{***}$ & $ (0.383)$ \\ 
 n\_languages & $0.666^{***}$ & $  (0.035)$ \\ 
 Constant & $-1.371^{***}$ & $(0.385)$ \\ 
\hline \\[-1.8ex] 
Observations & \multicolumn{1}{c}{26,281}\\
\hline \\[-1.8ex] 
\textit{Note:}  & \multicolumn{1}{r}{$^{*}$p$<$0.1; $^{**}$p$<$0.05; $^{***}$p$<$0.01} \\ 
\end{tabular} 
  \caption{Results of L.R. where the D.V. is whether user interacted on Twitter (mentions) and the I.V.s are conference $year$ and number of languages spoken.}  
  \vspace{-5mm}
  \label{tab-logit-results}
\end{table} 

\vspace{-2mm}  
\section{Related Work}
\label{section:related work}

There are several studies on the role of Twitter in academic conferences. Letierce \emph{et al.}~\cite{dyl43,Letierce2010} showed that Twitter is frequently used to spread information across researchers using the official conference hashtags. Wen \emph{et al.}~\cite{WenCSCW14} studied conference participants and found that newcomer students receive little attention from senior members of the research community. In an extension of this work, Wen \emph{et al.}~\cite{WenHT14} expand their research by analyzing 16 conferences over five years, identifying factors that contribute to the continuing participation of users to the online Twitter conference activity. We have continued this line of research by exploring the influence of language during conferences.
The role of language in Twitter has also been studied. Hong \emph{et al.} \cite{Lichan201} studied differences in usage patterns between language communities in Twitter, while Kim et al. \cite{KimHT14} performed a sociolinguistic study on the role of mono- and bilinguals in Twitter across multilingual societies such as Qatar, Quebec and Switzerland. Inspired by them, we adopt similar methods to build language communities but we target different lingua groups interacting at conferences.  

A broader but certainly related topic of study is the impact of \emph{culture} in online communication. Garcia \emph{et al.} ~\cite{GarciaGavilanes2014} studied the most discriminative features influencing international active conversation and attention in Twitter by mapping \emph{nationality} to I.P addresses (e-mails) or geolocated tweets. Language and nationality are two important cultural dimensions in people's identities, but we find that focusing on language(s) we capture the multicultural nature of most researchers that attend international conferences.



\vspace{-2mm} 
\section{Conclusions \& Future Work} \label{conclusions}

In this paper we show that the majority of users in Computer and Information Science conferences tweet only in English and most of the tweets are also posted in English. Nevertheless, our results indicate that members from other lingua communities produce most of the tweets and are more engaged than English monolinguals. 

A second observation is that although English is the lingua franca in academic conferences, apparently English monolinguals still prefer to interact more with themselves. The same happens for other important communities such as English-Japanese bilinguals. This is not the case for most of other important communities, who tend to interact more equally with members of other lingua. 

Our final finding is that there is more language diversity among people who interact with others on Twitter during conferences, compared to people who do not. This result suggests an important implication, which is that although English is the standard for scientific communication, the diversity in language use is a catalyst for interactions in a community.  

These findings leave us with several questions and encourage us to complement our work in several aspects. For example, which other aspects of people's culture can influence the communication gap across lingua groups? Can we identify that a research community requires more diversity by analyzing user interaction on Twitter? Can we identify user behavior related to specific lingua groups, such that we can differentiate English-Spanish bilinguals from English-German ones?  

\textbf{Acknowledgments:}
This work was carried out during the tenure of an ERCIM ``Alain Bensoussan'' fellowship program by the 4th author.




\small
\balance
\bibliographystyle{abbrv}
\bibliography{cultureht15,strings}
\normalsize
\appendix
\label{section-appendix}
The following tables show detailed data used in our analyses.

\begin{table*}[t!]
\centering
  \scalebox{0.8}{ 
\begin{tabular}{l|r|r|r|r|r|r|r|r|r|r|r|}
\cline{2-12}
 & \multicolumn{1}{c|}{en} & \multicolumn{1}{c|}{es} & \multicolumn{1}{c|}{fr} & \multicolumn{1}{c|}{pt} & \multicolumn{1}{c|}{ja} & \multicolumn{1}{c|}{ar} & \multicolumn{1}{c|}{nl} & \multicolumn{1}{c|}{ko} & \multicolumn{1}{c|}{it} & \multicolumn{1}{c|}{de} & \multicolumn{1}{c|}{total} \\ \hline
\multicolumn{1}{|l|}{AAAI} & 93\% & 1\% & 1\% & 0\% & 0\% & 0\% & 1\% & 0\% & 0\% & 0\% & 39,988 \\ \hline
\multicolumn{1}{|l|}{ACMMM} & 70\% & 4\% & 3\% & 1\% & 4\% & 0\% & 2\% & 1\% & 3\% & 4\% & 36,550 \\ \hline
\multicolumn{1}{|l|}{CHI} & 84\% & 3\% & 2\% & 1\% & 2\% & 0\% & 1\% & 1\% & 1\% & 1\% & 1,352,685 \\ \hline
\multicolumn{1}{|l|}{CIKM} & 80\% & 4\% & 1\% & 0\% & 2\% & 0\% & 1\% & 1\% & 0\% & 6\% & 239,461 \\ \hline
\multicolumn{1}{|l|}{ECIR} & 76\% & 8\% & 0\% & 1\% & 1\% & 1\% & 1\% & 0\% & 1\% & 1\% & 41,747 \\ \hline
\multicolumn{1}{|l|}{ECIS} & 75\% & 9\% & 1\% & 0\% & 0\% & 0\% & 7\% & 0\% & 0\% & 4\% & 27,466 \\ \hline
\multicolumn{1}{|l|}{HT} & 59\% & 3\% & 0\% & 3\% & 0\% & 0\% & 24\% & 0\% & 0\% & 2\% & 12,6873 \\ \hline
\multicolumn{1}{|l|}{ICIS} & 71\% & 0\% & 1\% & 0\% & 5\% & 11\% & 2\% & 0\% & 4\% & 2\% & 29,654 \\ \hline
\multicolumn{1}{|l|}{ICML} & 90\% & 3\% & 1\% & 0\% & 1\% & 0\% & 0\% & 1\% & 0\% & 1\% & 50,081 \\ \hline
\multicolumn{1}{|l|}{ICMT} & 92\% & 5\% & 1\% & 0\% & 0\% & 0\% & 0\% & 0\% & 1\% & 0\% & 27,445 \\ \hline
\multicolumn{1}{|l|}{ICSE} & 78\% & 6\% & 6\% & 3\% & 2\% & 0\% & 1\% & 0\% & 1\% & 1\% & 227,072 \\ \hline
\multicolumn{1}{|l|}{ISMAR} & 80\% & 1\% & 1\% & 0\% & 7\% & 0\% & 2\% & 3\% & 0\% & 2\% & 61,103 \\ \hline
\multicolumn{1}{|l|}{IUI} & 81\% & 4\% & 1\% & 0\% & 0\% & 0\% & 7\% & 1\% & 0\% & 4\% & 36,028 \\ \hline
\multicolumn{1}{|l|}{KDD} & 88\% & 3\% & 1\% & 1\% & 1\% & 1\% & 1\% & 0\% & 1\% & 0\% & 157,607 \\ \hline
\multicolumn{1}{|l|}{MobileHCI} & 76\% & 5\% & 1\% & 0\% & 0\% & 0\% & 2\% & 2\% & 1\% & 3\% & 25,572 \\ \hline
\multicolumn{1}{|l|}{NIPS} & 81\% & 2\% & 5\% & 0\% & 5\% & 0\% & 0\% & 2\% & 0\% & 1\% & 161,394 \\ \hline
\multicolumn{1}{|l|}{SIGGRAPH} & 84\% & 6\% & 2\% & 1\% & 1\% & 1\% & 0\% & 1\% & 0\% & 0\% & 1,096,484 \\ \hline
\multicolumn{1}{|l|}{SIGIR} & 86\% & 2\% & 0\% & 1\% & 1\% & 2\% & 0\% & 1\% & 0\% & 1\% & 138,094 \\ \hline
\multicolumn{1}{|l|}{SIGMOD} & 79\% & 1\% & 1\% & 2\% & 4\% & 1\% & 0\% & 5\% & 0\% & 1\% & 38,759 \\ \hline
\multicolumn{1}{|l|}{SLE} & 73\% & 2\% & 5\% & 1\% & 1\% & 0\% & 14\% & 0\% & 0\% & 2\% & 39,885 \\ \hline
\multicolumn{1}{|l|}{UBICOMP} & 79\% & 4\% & 1\% & 0\% & 6\% & 0\% & 1\% & 2\% & 1\% & 4\% & 75,100 \\ \hline
\multicolumn{1}{|l|}{UIST} & 77\% & 1\% & 1\% & 0\% & 17\% & 0\% & 0\% & 0\% & 0\% & 1\% & 51,563 \\ \hline
\multicolumn{1}{|l|}{VLDB} & 86\% & 2\% & 0\% & 0\% & 3\% & 1\% & 0\% & 0\% & 2\% & 1\% & 47,701 \\ \hline
\multicolumn{1}{|l|}{WSDM} & 85\% & 4\% & 1\% & 0\% & 1\% & 0\% & 0\% & 0\% & 1\% & 1\% & 53,951 \\ \hline
\multicolumn{1}{|l|}{WWW} & 64\% & 5\% & 18\% & 6\% & 0\% & 0\% & 1\% & 0\% & 1\% & 1\% & 1,785,006 \\ \hline
\multicolumn{1}{|l|}{XP} & 79\% & 9\% & 1\% & 2\% & 0\% & 0\% & 1\% & 0\% & 1\% & 3\% & 231,959 \\ \hline
\end{tabular}
}
\vspace{-3mm}
\caption{The number of tweets per conference and the percentage of tweets in each language.}
\label{tbl-languages-conferences-top}
\vspace{-3mm}
\end{table*}

\begin{table*}[h!]
\centering
\resizebox{\textwidth}{!}{%
\begin{tabular}{l|l|r|r|r|r|r|r|r|r|r|r|r|r|r|r|r|r|r|r|r|r|r|r|r|r|r|r|}
\cline{2-28}
 & & \multicolumn{1}{l|}{AAAI} & \multicolumn{1}{l|}{ACMMM} & \multicolumn{1}{l|}{CHI} & \multicolumn{1}{l|}{CIKM} & \multicolumn{1}{l|}{ECIR} & \multicolumn{1}{l|}{ECIS} & \multicolumn{1}{l|}{HT} & \multicolumn{1}{l|}{ICIS} & \multicolumn{1}{l|}{ICML} & \multicolumn{1}{l|}{ICMT} & \multicolumn{1}{l|}{ICSE} & \multicolumn{1}{l|}{ISMAR} & \multicolumn{1}{l|}{IUI} & \multicolumn{1}{l|}{KDD} & \multicolumn{1}{l|}{MobileHCI} & \multicolumn{1}{l|}{NIPS} & \multicolumn{1}{l|}{SIGGRAPH} & \multicolumn{1}{l|}{SIGIR} & \multicolumn{1}{l|}{SIGMOD} & \multicolumn{1}{l|}{SLE} & \multicolumn{1}{l|}{UBICOMP} & \multicolumn{1}{l|}{UIST} & \multicolumn{1}{l|}{VLDB} & \multicolumn{1}{l|}{WSDM} & \multicolumn{1}{l|}{WWW} & \multicolumn{1}{l|}{XP} \\ \hline
\multicolumn{1}{|l|}{\multirow{4}{*}{2009}} & tweets & 0 & 22 & 38932 & 122 & 40 & 2666 & 2559 & 87 & 12 & 3330 & 306 & 11746 & 176 & 619 & 543 & 1279 & 7932 & 27176 & 2 & 1719 & 241 & 1359 & 679 & 0 & 77243 & 2111 \\ \cline{2-28} 
\multicolumn{1}{|l|}{} & retweets & 0 & 1 & 459 & 14 & 11 & 30 & 169 & 8 & 0 & 3 & 14 & 327 & 2 & 355 & 32 & 44 & 213 & 745 & 1 & 151 & 46 & 107 & 139 & 1 & 770 & 263 \\ \cline{2-28} 
\multicolumn{1}{|l|}{} & mentions & 0 & 2 & 3932 & 74 & 34 & 64 & 719 & 27 & 4 & 19 & 76 & 881 & 41 & 498 & 81 & 293 & 954 & 2040 & 3 & 549 & 192 & 540 & 412 & 16 & 3753 & 1701 \\ \cline{2-28} 
\multicolumn{1}{|l|}{} & urls & 0 & 4 & 24102 & 33 & 26 & 2132 & 1111 & 35 & 1 & 1275 & 54 & 8003 & 21 & 330 & 100 & 724 & 2329 & 16639 & 0 & 1109 & 105 & 637 & 241 & 0 & 42538 & 364 \\ \hline
\multicolumn{1}{|l|}{\multirow{4}{*}{2010}} & tweets & 324 & 6822 & 152637 & 2753 & 4552 & 91 & 13837 & 700 & 91 & 343 & 15641 & 2054 & 4271 & 6586 & 747 & 18196 & 7916 & 3360 & 2448 & 4931 & 4912 & 2306 & 1120 & 5450 & 196200 & 25610 \\ \cline{2-28} 
\multicolumn{1}{|l|}{} & retweets & 93 & 935 & 9613 & 567 & 540 & 9 & 578 & 53 & 8 & 128 & 441 & 100 & 746 & 255 & 256 & 1056 & 239 & 506 & 415 & 371 & 983 & 358 & 201 & 543 & 10203 & 3142 \\ \cline{2-28} 
\multicolumn{1}{|l|}{} & mentions & 223 & 1798 & 35212 & 1119 & 1379 & 51 & 1036 & 139 & 18 & 289 & 1016 & 259 & 1590 & 582 & 489 & 1877 & 633 & 1321 & 896 & 693 & 1929 & 994 & 564 & 1532 & 22306 & 9670 \\ \cline{2-28} 
\multicolumn{1}{|l|}{} & urls & 72 & 3117 & 96384 & 826 & 3540 & 30 & 11039 & 276 & 8 & 113 & 11152 & 1040 & 3172 & 2330 & 363 & 9276 & 2505 & 961 & 910 & 2442 & 2658 & 1325 & 597 & 3157 & 124629 & 6737 \\ \hline
\multicolumn{1}{|l|}{\multirow{4}{*}{2011}} & tweets & 14514 & 2449 & 204498 & 9091 & 5440 & 1297 & 15815 & 2634 & 3628 & 4364 & 19659 & 7013 & 4436 & 16312 & 3263 & 11182 & 46757 & 4577 & 3298 & 847 & 10205 & 2586 & 5447 & 5668 & 117833 & 44334 \\ \cline{2-28} 
\multicolumn{1}{|l|}{} & retweets & 243 & 488 & 13411 & 2401 & 924 & 109 & 917 & 53 & 557 & 564 & 2403 & 223 & 911 & 1423 & 271 & 1376 & 1782 & 1240 & 616 & 2 & 1162 & 724 & 744 & 1308 & 6478 & 6971 \\ \cline{2-28} 
\multicolumn{1}{|l|}{} & mentions & 533 & 982 & 33244 & 4845 & 187a0 & 280 & 2343 & 195 & 1154 & 1261 & 4683 & 438 & 1709 & 2942 & 967 & 2690 & 3674 & 2822 & 1116 & 41 & 2856 & 1773 & 1960 & 2068 & 12703 & 22277 \\ \cline{2-28} 
\multicolumn{1}{|l|}{} & urls & 3744 & 1155 & 103827 & 4459 & 2190 & 599 & 4335 & 645 & 1291 & 962 & 8217 & 5856 & 1718 & 8702 & 1106 & 6408 & 20819 & 1879 & 1615 & 181 & 5604 & 950 & 2177 & 3087 & 71491 & 13372 \\ \hline
\multicolumn{1}{|l|}{\multirow{4}{*}{2012}} & tweets & 1182 & 5229 & 102553 & 14859 & 4607 & 4921 & 13425 & 1953 & 5836 & 781 & 18219 & 7639 & 4084 & 11632 & 5199 & 15283 & 178229 & 17235 & 7987 & 4459 & 3738 & 8186 & 6729 & 11780 & 351146 & 18580 \\ \cline{2-28} 
\multicolumn{1}{|l|}{} & retweets & 120 & 941 & 6580 & 2628 & 1085 & 106 & 1855 & 125 & 1183 & 149 & 2881 & 530 & 570 & 1173 & 750 & 1552 & 5714 & 2345 & 967 & 689 & 458 & 611 & 947 & 1595 & 19521 & 3604 \\ \cline{2-28} 
\multicolumn{1}{|l|}{} & mentions & 350 & 1884 & 18383 & 5342 & 2098 & 209 & 3647 & 228 & 2801 & 410 & 5922 & 1070 & 1134 & 1727 & 1538 & 3668 & 12588 & 4477 & 2092 & 1739 & 1060 & 1431 & 1916 & 3569 & 46939 & 11195 \\ \cline{2-28} 
\multicolumn{1}{|l|}{} & urls & 530 & 2511 & 46343 & 6953 & 2437 & 2880 & 5567 & 945 & 2529 & 205 & 7280 & 6700 & 1693 & 7904 & 2544 & 8755 & 102146 & 7400 & 3282 & 1691 & 1880 & 6125 & 3303 & 5338 & 153339 & 5875 \\ \hline
\multicolumn{1}{|l|}{\multirow{4}{*}{2013}} & tweets & 3297 & 5385 & 183168 & 96779 & 5679 & 3712 & 13961 & 9197 & 14584 & 4376 & 57411 & 2560 & 3681 & 44026 & 2512 & 39327 & 306407 & 18409 & 5580 & 5048 & 20140 & 9867 & 9501 & 4939 & 144870 & 22022 \\ \cline{2-28} 
\multicolumn{1}{|l|}{} & retweets & 203 & 1790 & 8436 & 5613 & 1078 & 206 & 994 & 527 & 1489 & 337 & 3566 & 268 & 732 & 3851 & 233 & 3409 & 10726 & 2798 & 754 & 730 & 913 & 1146 & 738 & 1265 & 9339 & 2972 \\ \cline{2-28} 
\multicolumn{1}{|l|}{} & mentions & 499 & 3648 & 23925 & 13310 & 2707 & 364 & 2953 & 871 & 3791 & 1135 & 7714 & 629 & 1863 & 7457 & 713 & 8125 & 23287 & 6126 & \multicolumn{1}{l|}{1533} & 1935 & 2176 & 2234 & 1666 & 2416 & 21264 & 9962 \\ \cline{2-28} 
\multicolumn{1}{|l|}{} & urls & 1027 & 2972 & 84575 & 44217 & 3166 & 2504 & 6030 & 5807 & 7587 & 914 & 22559 & 1359 & 1739 & 22637 & 1391 & 20743 & 178613 & 8990 & \multicolumn{1}{l|}{2461} & 1557 & 14087 & 6012 & 5585 & 2430 & 66205 & 7484 \\ \hline
\end{tabular}
}
\vspace{-3mm}
\caption{Metrics of tweets, retweets, mentions and tweets with URLs per conference and year.}
\vspace{-3mm}
\label{tbl-metrics}
\end{table*}

\begin{table*}[h!]
\centering
\resizebox{\textwidth}{!}{%
\begin{tabular}{l|l|r|l|r|l|r|l|r|l|r|l|r|l|r|l|r|l|r|l|l|l|r|l|r|l|r|l|r|l|r|l|r|l|r|l|r|l|r|l|r|l|r|l|r|l|r|l|r|l|r|l|r|}
\cline{2-53}
 & \multicolumn{2}{l|}{AAAI} & \multicolumn{2}{l|}{ACMMM} & \multicolumn{2}{l|}{CHI} & \multicolumn{2}{l|}{CIKM} & \multicolumn{2}{l|}{ECIR} & \multicolumn{2}{l|}{ECIS} & \multicolumn{2}{l|}{HT} & \multicolumn{2}{l|}{ICIS} & \multicolumn{2}{l|}{ICML} & \multicolumn{2}{l|}{ICMT} & \multicolumn{2}{l|}{ICSE} & \multicolumn{2}{l|}{ISMAR} & \multicolumn{2}{l|}{IUI} & \multicolumn{2}{l|}{KDD} & \multicolumn{2}{l|}{MobileHCI} & \multicolumn{2}{l|}{NIPS} & \multicolumn{2}{l|}{SIGGRAPH} & \multicolumn{2}{l|}{SIGIR} & \multicolumn{2}{l|}{SIGMOD} & \multicolumn{2}{l|}{SLE} & \multicolumn{2}{l|}{UBICOMP} & \multicolumn{2}{l|}{UIST} & \multicolumn{2}{l|}{VLDB} & \multicolumn{2}{l|}{WSDM} & \multicolumn{2}{l|}{WWW} & \multicolumn{2}{l|}{XP} \\ \hline
\multicolumn{1}{|l|}{\multirow{5}{*}{2009}} &  &  & en & 4 & en & 335 & en & 8 & en & 5 & en & 8 & en & 43 & en & 10 & en & 1 & en & 3 & en & 3 & en & 66 & en & 2 & en & 14 & en & 10 & en & 8 & en & 66 & en & 97 & en & 1 & en & 32 & en & 21 & en & 41 & en & 21 & en & 1 & en & 313 & en & 36 \\ \cline{2-53} 
\multicolumn{1}{|l|}{} &  &  & de & 2 & it & 35 & de & 2 & es & 1 & de & 4 & it & 10 & de & 1 &  &  & ar & 1 & fr & 1 & es & 12 &  &  & es & 2 & de & 4 & de & 2 & fr & 15 & es & 12 &  &  & de & 8 & ja & 4 & ja & 6 & fr & 5 & es & 1 & es & 112 & no & 9 \\ \cline{2-53} 
\multicolumn{1}{|l|}{} &  &  & fr & 1 & ps & 25 & ja & 1 & hu & 1 & es & 3 & pt & 10 & fi & 1 &  &  & de & 1 & hi & 1 & de & 10 &  &  & fr & 2 & it & 2 & es & 2 & it & 15 & it & 9 &  &  & es & 5 & fr & 2 & de & 3 & it & 3 & fr & 1 & it & 77 & sv & 6 \\ \cline{2-53} 
\multicolumn{1}{|l|}{} &  &  & nl & 1 & es & 23 & nl & 1 & is & 1 & et & 2 & es & 7 & it & 1 &  &  & es & 1 & ur & 1 & fr & 10 &  &  & ar & 1 & nl & 2 & fr & 2 & es & 12 & no & 9 &  &  & no & 3 & nl & 2 & ar & 2 & ja & 2 & it & 1 & ca & 60 & it & 5 \\ \cline{2-53} 
\multicolumn{1}{|l|}{} &  &  & ro & 1 & de & 22 & no & 1 & it & 1 & fi & 2 & de & 6 & ko & 1 &  &  & et & 1 &  &  & nl & 9 &  &  & el & 1 & ps & 2 & it & 2 & ps & 7 & pt & 8 &  &  & pt & 3 & cs & 1 & es & 2 & ro & 2 & ps & 1 & de & 58 & es & 4 \\ \hline
\multicolumn{1}{|l|}{\multirow{5}{*}{2010}} & en & 20 & en & 100 & en & 802 & en & 54 & en & 58 & en & 7 & en & 54 & en & 35 & en & 4 & en & 20 & en & 87 & en & 39 & en & 48 & en & 49 & en & 17 & en & 132 & en & 71 & en & 91 & en & 37 & en & 41 & en & 87 & en & 53 & en & 19 & en & 77 & en & 1140 & en & 470 \\ \cline{2-53} 
\multicolumn{1}{|l|}{} & es & 4 & it & 30 & fr & 88 & it & 6 & fr & 9 & de & 2 & de & 13 & de & 3 & de & 1 & es & 6 & de & 14 & ko & 12 & de & 10 & es & 8 & ca & 4 & fr & 19 & it & 12 & es & 13 & de & 3 & de & 12 & it & 15 & ja & 14 & it & 3 & es & 17 & es & 263 & no & 143 \\ \cline{2-53} 
\multicolumn{1}{|l|}{} & it & 4 & fr & 15 & de & 87 & es & 5 & it & 8 & nl & 2 & it & 11 & ja & 3 &  &  & it & 5 & it & 14 & nl & 10 & it & 9 & it & 7 & de & 3 & ja & 15 & de & 11 & it & 8 & es & 3 & nl & 11 & es & 11 & de & 5 & ro & 3 & fr & 7 & it & 221 & es & 59 \\ \cline{2-53} 
\multicolumn{1}{|l|}{} & ca & 2 & es & 14 & it & 83 & fr & 4 & de & 7 & no & 2 & es & 8 & no & 3 &  &  & nl & 3 & es & 12 & es & 8 & es & 8 & fr & 6 & es & 3 & de & 14 & es & 11 & ca & 7 & fr & 3 & es & 7 & de & 8 & nl & 5 & es & 2 & it & 7 & fr & 172 & de & 53 \\ \cline{2-53} 
\multicolumn{1}{|l|}{} & de & 2 & nl & 11 & es & 81 & ca & 3 & es & 7 & sv & 2 & no & 7 & ar & 2 &  &  & ca & 2 & pt & 11 & de & 6 & ko & 6 & ar & 5 & it & 2 & es & 14 & no & 11 & nl & 7 & it & 3 & fr & 7 & ja & 7 & fr & 3 & fr & 2 & pt & 7 & de & 166 & it & 50 \\ \hline
\multicolumn{1}{|l|}{\multirow{5}{*}{2011}} & en & 64 & en & 75 & en & 1280 & en & 157 & en & 98 & en & 44 & en & 141 & en & 30 & en & 52 & en & 30 & en & 270 & en & 44 & en & 62 & en & 189 & en & 77 & en & 185 & en & 293 & en & 88 & en & 63 & fr & 6 & en & 80 & en & 69 & en & 103 & en & 86 & en & 630 & en & 538 \\ \cline{2-53} 
\multicolumn{1}{|l|}{} & de & 9 & es & 16 & fr & 201 & es & 25 & es & 22 & de & 10 & nl & 93 & de & 6 & es & 8 & es & 10 & es & 62 & it & 12 & de & 16 & es & 30 & sv & 14 & es & 24 & fr & 61 & es & 17 & el & 10 & en & 4 & ja & 21 & es & 9 & ja & 15 & es & 17 & es & 165 & es & 189 \\ \cline{2-53} 
\multicolumn{1}{|l|}{} & it & 9 & it & 9 & es & 180 & de & 17 & de & 12 & nl & 8 & es & 39 & no & 6 & it & 6 & fr & 10 & it & 54 & de & 8 & es & 8 & it & 28 & de & 11 & fr & 23 & it & 55 & it & 16 & de & 6 & es & 4 & de & 15 & fr & 8 & de & 14 & de & 10 & it & 147 & it & 104 \\ \cline{2-53} 
\multicolumn{1}{|l|}{} & tr & 9 & de & 7 & it & 162 & it & 15 & it & 11 & fi & 7 & de & 36 & es & 3 & fr & 5 & it & 6 & pt & 53 & fr & 8 & it & 7 & de & 19 & no & 9 & it & 21 & es & 53 & de & 11 & ko & 6 & it & 4 & it & 11 & de & 5 & it & 14 & it & 7 & fr & 122 & ca & 85 \\ \cline{2-53} 
\multicolumn{1}{|l|}{} & es & 8 & ja & 7 & ps & 132 & ca & 12 & no & 8 & it & 5 & no & 33 & it & 3 & ro & 5 & da & 5 & ro & 35 & es & 7 & nl & 7 & fr & 19 & it & 8 & de & 17 & no & 38 & ja & 8 & es & 4 & ca & 3 & fr & 9 & fa & 4 & es & 8 & zh-hans & 7 & de & 111 & no & 85 \\ \hline
\multicolumn{1}{|l|}{\multirow{5}{*}{2012}} & en & 37 & en & 101 & en & 1218 & en & 207 & en & 96 & en & 58 & en & 145 & en & 46 & en & 109 & en & 13 & en & 277 & en & 46 & en & 41 & en & 82 & en & 82 & en & 230 & en & 1216 & en & 283 & en & 119 & en & 71 & en & 90 & en & 86 & en & 117 & en & 190 & en & 2616 & en & 216 \\ \cline{2-53} 
\multicolumn{1}{|l|}{} & es & 7 & ja & 23 & es & 151 & es & 37 & es & 25 & es & 22 & es & 29 & de & 8 & de & 11 & es & 4 & es & 47 & ja & 30 & de & 9 & zh-hans & 19 & es & 12 & es & 27 & es & 243 & es & 36 & it & 18 & nl & 34 & it & 11 & ja & 11 & es & 15 & es & 23 & fr & 1625 & no & 41 \\ \cline{2-53} 
\multicolumn{1}{|l|}{} & nl & 4 & fr & 16 & fr & 127 & it & 25 & ca & 14 & nl & 14 & it & 24 & it & 6 & es & 11 & nl & 4 & it & 47 & es & 6 & es & 8 & de & 10 & de & 11 & fr & 18 & it & 198 & it & 32 & es & 15 & de & 16 & fr & 9 & es & 7 & it & 13 & it & 20 & es & 994 & de & 40 \\ \cline{2-53} 
\multicolumn{1}{|l|}{} & pt & 3 & it & 16 & it & 113 & de & 22 & it & 14 & ca & 11 & de & 23 & nl & 6 & fr & 8 & fr & 3 & pt & 44 & zh-hans & 6 & it & 7 & zh-hant & 9 & it & 11 & it & 18 & fr & 182 & de & 23 & de & 11 & es & 11 & de & 7 & de & 5 & tr & 11 & de & 17 & it & 821 & sv & 40 \\ \cline{2-53} 
\multicolumn{1}{|l|}{} & ro & 3 & de & 11 & de & 107 & fr & 22 & de & 9 & it & 11 & nl & 23 & es & 5 & ru & 7 & de & 2 & de & 42 & fr & 5 & ca & 6 & fr & 8 & fr & 10 & ja & 16 & de & 159 & no & 17 & fr & 10 & it & 9 & ca & 6 & fr & 5 & de & 10 & fr & 17 & ca & 627 & es & 39 \\ \hline
\multicolumn{1}{|l|}{\multirow{5}{*}{2013}} & en & 48 & en & 159 & en & 1605 & en & 799 & en & 124 & en & 89 & nl & 278 & en & 135 & en & 149 & en & 25 & en & 425 & en & 55 & en & 71 & en & 472 & en & 70 & en & 518 & en & 2281 & en & 378 & en & 111 & en & 66 & en & 158 & en & 222 & en & 115 & en & 96 & en & 1158 & en & 284 \\ \cline{2-53} 
\multicolumn{1}{|l|}{} & es & 10 & es & 34 & fr & 343 & es & 131 & ru & 25 & nl & 28 & en & 260 & it & 25 & es & 15 & es & 8 & es & 72 & fr & 7 & es & 13 & es & 82 & de & 11 & ja & 57 & es & 433 & it & 40 & fr & 13 & fr & 13 & de & 26 & ja & 81 & it & 29 & es & 20 & es & 412 & de & 77 \\ \cline{2-53} 
\multicolumn{1}{|l|}{} & de & 6 & fr & 29 & es & 232 & it & 110 & bg & 18 & de & 19 & no & 61 & nl & 19 & ja & 14 & fr & 6 & it & 67 & no & 6 & de & 9 & it & 52 & es & 11 & fr & 54 & it & 372 & es & 39 & ro & 11 & de & 11 & es & 15 & it & 16 & es & 16 & it & 18 & pt & 368 & it & 59 \\ \cline{2-53} 
\multicolumn{1}{|l|}{} & it & 6 & ca & 19 & it & 190 & fr & 109 & uk & 18 & fr & 8 & de & 44 & de & 17 & de & 13 & hu & 6 & fr & 64 & it & 5 & fr & 8 & fr & 40 & it & 6 & it & 51 & fr & 358 & de & 28 & de & 9 & es & 10 & ja & 14 & de & 15 & de & 13 & fr & 11 & it & 295 & es & 37 \\ \cline{2-53} 
\multicolumn{1}{|l|}{} & ps & 6 & de & 15 & de & 162 & de & 107 & sr & 17 & sv & 7 & es & 28 & sv & 15 & fr & 13 & it & 5 & de & 55 & ja & 4 & it & 7 & de & 38 & ca & 5 & es & 50 & de & 247 & no & 26 & es & 8 & nl & 10 & nl & 12 & es & 15 & no & 11 & de & 8 & fr & 237 & sv & 33 \\ \hline
\end{tabular}
}
\vspace{-3mm}
\caption{Top three languages at every conference each year based on the number of users tweeting in each language.}
\label{tbl-languages-top}
\vspace{-3mm}
\end{table*}

\end{document}